\documentclass{amsart}
\usepackage{amsmath,amssymb,amsfonts, bbm}
\usepackage{hyperref}
\usepackage[T1]{fontenc}
\usepackage[pdftex]{graphicx}

\newcommand{\CC}{\mathbb C}
\newcommand{\U}{\mathcal{U}}
\renewcommand{\u}{\mathfrak{u}}
\newcommand{\I}{\mathbbm{I}}
\newcommand{\1}{{\mathbf 1}}
\newcommand{\0}{{\mathbf 0}}
\newcommand{\ptexp}{\operatorname{exp_{+}\!}}
\newcommand{\diag}{\operatorname{diag}}
\newcommand{\T}{\operatorname{T}\!}
\renewcommand{\H}{\operatorname{H}\!}
\newcommand{\V}{\operatorname{V}\!}
\newcommand{\HH}{\mathcal{H}}
\newcommand{\J}{\mathcal{J}}
\newcommand{\D}{\mathcal{D}}
\newcommand{\A}{\mathcal{A}}
\renewcommand{\S}{\mathcal{S}}
\newcommand{\KK}{\mathcal{K}}
\renewcommand{\th}{\text{th}}

\newcommand{\dt}{\operatorname{d}\!t}
\newcommand{\Hol}{\operatorname{Hol}}
\newcommand{\ket}[1]{|{#1}\rangle}

\newcommand{\braket}[2]{\langle #1|#2\rangle}
\newcommand{\ketbra}[2]{|#1\rangle\langle #2|}
\newcommand{\geop}{\gamma_{\text{geo}}}
\newcommand{\geopf}{\zeta_{\text{geo}}}
\newcommand{\tr}{\operatorname{Tr}}
\begin{document}
\title[Geometric phase for mixed states]{Operational geometric phase for mixed quantum states}
\date{\today}
\author{Ole Andersson and Hoshang Heydari}
\address{Department of Physics, Stockholm University, 10691 Stockholm, Sweden}
\email{olehandersson@gmail.com, hoshang@fysik.su.se}
\maketitle
\begin{abstract}
Geometric phase has found a broad spectrum of applications in both classical and quantum physics, such as condensed matter and quantum computation.
In this paper we introduce an operational geometric phase for mixed quantum states, based on spectral weighted traces of holonomies, and we prove that it generalizes the standard definition of geometric phase for mixed states, which is based on quantum interferometry.
We also introduce higher order geometric phases, and prove that under a fairly weak, generically satisfied, requirement, there is always a well-defined geometric phase of some order. Our approach applies to general unitary evolutions of both non\-degenerate and degenerate mixed states. 
Moreover, since we provide an explicit formula for the geometric phase that can be easily implemented, it is particularly well suited for computations in quantum physics.
\end{abstract}

\maketitle

\section{Introduction}
In the seminal paper \cite{Berry1984}, Berry demonstrated that a quantum system in a pure state that undergoes an adiabatic cyclic evolution can pick up a phase which is independent of the dynamics of the system.
Berry's publication began a, still ongoing, era of intensive research on holonomy effects in quantum theory, and some of the more important earlier papers on the topic have been reprinted in \cite{Shapere_etal1989}.
There, for example, one can find \cite{Simon1983}, in which Simon identifies Berry's phase as a parallel transport holonomy in a Hermitian line bundle over the parameter manifold of the system's Hamiltonian, and \cite{Aharonov_etal1987},          
in which Aharonov and Anandan extend Berry's result to the non-adiabatic cyclic case, and make explicit that Berry's phase is due to the curvature of projective Hilbert space.
More recent applications of quantum mechanics in which holonomy effects play a crucial role include quantum information processing and quantum computing \cite{Pachos_etal1990, Zanardi_etal1999, Ekert_etal2000, Zanardi_etal2007, Rezakhani_etal2010, Jones_etal2000, Falci_etal2000, Farhi_etal2001, Duan_etal2001, Recati_etal2002}. 
Indeed, it is believed that quantum computation implemented by holonomic quantum gates is intrinsically fault-tolerant and robust against noise \cite{Nazir_etal2002, Solinas_etal2003, Zhu_etal2005}.

States of experimentally prepared quantum systems generally 
exhibit classical uncertainty.
Hence they are most appropriately described as mixtures of pure states. 
Uhlmann \cite{Uhlmann1986, Uhlmann1991} was among the first to 
develop a theory for geometric phase for parallel transported mixed states.
The theory is based on the concept of purification.
Another approach to geometric phase for mixed states, based on quantum interferometry, was proposed by Sj\"{o}qvist \emph{et al.} \cite{Sjoqvist_etal2000}.
They derived, and carefully analyzed, an expression for the intensity of the output signal in a classical Mach-Zehnder interferometer, and from this they extracted a definition of geometric phase for mixed states.
Several experiments \cite{Du_etal2003, Ericsson_etal2005, Klepp_etal2005, Ghosh_etal2006} have been carried out that verify Sj\"{o}qvist \emph{et al.}'s results.

A shortcoming of the definition of geometric phase given in \cite{Sjoqvist_etal2000} is that it only applies to parallel transported non\-degenerate mixed states.
In the current paper we provide a geometrical framework, inspired by arguments of Montgomery \cite{Montgomery1991}, in which geometric phase for unitarily evolving general mixed states has a natural definition.
We show that this definition agrees with the definition in \cite{Sjoqvist_etal2000} in those cases where their definition applies, and we derive an explicit formula for the geometric phase which can be easily implemented. 
Our approach is thus particularly well suited for computations. 

The geometric phase for an evolving quantum system prepared in a pure state is undefined if the initial and final states are orthogonal. 
For quantum systems in mixed states a similar situation may occur. In a concluding section we introduce higher order geometric phases, and
prove that under a generically satisfied condition there will always be a well-defined geometric phase of some order. We also relate the higher order geometric phases to the off-diagonal geometric phases studied by Manini and Pistolesi \cite{Manini_etal2000}, and Mukunda \emph{et al.} \cite{Mukunda_etal2001}.

\section{Bundles of purifications}
\subsection{Mixed states and purifications}
Mixed quantum states can be represented by density operators. 
A density operator is a self-adjoint, nonnegative, trace-class operator with unit trace.
We write $\D(\HH)$ for the space of density operators on a Hilbert space $\HH$, and
$\D_n(\HH)$ for the subspace of density operators with finite rank at most $n$.

A state is called pure if its density operator has rank $1$.
In quantum mechanics, especially quantum information theory, \emph{purification} refers to the fact that every density operator can be thought of as representing the reduced state of some pure state.
Indeed, one can show that if $\rho$ is a density operator acting on $\HH$, there exists a Hilbert space $\KK$ and a density operator $R$ representing a pure state in $\HH\otimes \KK$ such that $\rho$ is the partial trace of $R$ with respect to $\KK$. 
In this paper we only consider quantum systems whose states are represented by density operators with finite rank, and by the \emph{standard purification} of the density operators in $\D_n(\HH)$ we mean the surjective map 
\begin{equation}\label{st pur}
\pi:\S(\HH\otimes\CC^{n*})\to\D_n(\HH),\quad
\ket{\Psi}\mapsto\tr_{\CC^{n*}}\ketbra{\Psi}{\Psi}.
\end{equation}
Here $\CC^{n*}$ is the space of linear functionals on $\CC^n$, and $\S(\HH\otimes \CC^{n*})$ is the unit sphere in $\HH\otimes \CC^{n*}$. 

Evolving mixed states will be represented by piecewise smooth curves of density operators that, for convenience, are assumed to be defined on an unspecified interval $0\leq t\leq \tau$. Operators on $\CC^n$ will be represented by matrices with respect to the canonical basis, whose $k^\th$ member we denote by $e_k$. We write $\1_n$ and $\0_n$ for the $n\times n$ identity matrix and zero matrix, respectively.
 
\subsection{Bundles of purifications over orbits of isospectral density operators} 
When you purify a state you artificially add extraneous, perhaps inaccessible, information to your state.
One way to get rid of this extra information is to use symplectic reduction.
Montgomery \cite{Montgomery1991} showed that if one does symplectic reduction on $\HH\otimes\CC^{n*}$
by the action of the unitary group of $\CC^{n*}$,  then the reduced phase spaces can be identified with the (co)adjoint orbits of isospectral density operators in $\D_n(\HH)$, and the canonical reduced space submersions are given by certain spectral weighted eigenframe bundles. 
(A precise reformulation of Montgomery's result will be given below.) 
We show that in this context, there is a natural definition of geometric phase for unitarily evolving mixed states which generalizes the definition of Sj\"{o}qvist \emph{et al.}
But before we do that we need to translate Montgomery's result into a language appropriate for our purposes. 
We begin by identifying $\HH\otimes \CC^{n*}$ with the space of linear maps from $\CC^n$ to $\HH$, equipped with the Hilbert-Schmidt Hermitian inner product. 
Then the standard purification \eqref{st pur} is given by $\pi(\Psi)=\Psi\Psi^\dagger$.

A density operator whose evolution is governed by a von Neumann equation stays in a single orbit of the left conjugation action of the unitary group $\U(\HH)$ on $\D(\HH)$. 
These orbits are in one-to-one correspondence with the set of possible spectra for density operators on $\HH$, where by the spectrum of a rank $n$ density operator we mean the \emph{nonincreasing} sequence $\sigma =\left(p_1,p_2,\dots,p_n\right)$ of its, not necessarily distinct, \emph{positive} eigenvalues. Throughout we assume $\sigma$ to be fixed, and we write $\D(\sigma)$ for the corresponding orbit of density operators. Also, we refer to the integer $n$ as the length of $\sigma$.
 
Let $P(\sigma)$ be the diagonal $n\times n$ matrix with diagonal $\sigma$, and define $\pi:\S(\sigma)\to \D(\sigma)$
as the restriction of \eqref{st pur} to $\S(\sigma)=\{\Psi\in\S(\HH\otimes\CC^{n*}):\Psi^\dagger \Psi=P(\sigma)\}$.
Then $\pi$ is a principal fiber bundle with gauge group $\U(\sigma)=\{U\in\U(n):UP(\sigma)=P(\sigma)U\}$.
We equip $\S(\sigma)$ with the metric given by $2\hbar$ times the real part of the Hilbert-Schmidt Hermitian inner product:
$G(X,Y)=\hbar\tr(X^\dagger Y+Y^\dagger X)$.

The vertical and horizontal bundles of $\pi$ are the 
subbundles $\V\S(\sigma)$ and $\H\S(\sigma)$, respectively, of the tangent bundle $\T\S(\sigma)$.
The former consists of the kernels of 
the differential of $\pi$, and the latter
of the kernels' orthogonal complements with respect to $G$.
Vectors in $\V\S(\sigma)$ and $\H\S(\sigma)$
are called vertical and horizontal, respectively, and a curve in $\S(\sigma)$ is called horizontal if all of its velocity vectors are horizontal.
Recall that for every curve $\rho$ in $\D(\sigma)$ and every $\Psi_0$ in the fiber over $\rho(0)$ there is a unique horizontal curve $\Psi$ in $\S(\sigma)$ that starts at $\Psi_0$ and projects onto $\rho$. This curve is the \emph{horizontal lift} of $\rho$ extending from $\Psi_0$.
Furthermore, if $\rho$ happens to be a closed curve, then the \emph{holonomy of $\rho$ with base point $\Psi_0$} is the unique matrix $\Hol(\rho,\Psi_0)$ in $\U(\sigma)$ satisfying 
$\Psi(\tau)=\Psi_0\Hol(\rho,\Psi_0)$.
We define the geometric phase of $\rho$ to be the argument of the spectral weighted trace of $\Hol(\rho,\Psi_0)$:
\begin{equation}\label{phase from holonomy} 
\geop(\rho)=\arg\tr\left(P(\sigma)\Hol(\rho,\Psi_0)\right).
\end{equation}
Observe that \eqref{phase from holonomy} does not depend on $\Psi_0$ because the holonomy of $\rho$ associated with a different base point is conjugate equivalent to $\Hol(\rho,\Psi_0)$ in $\U(\sigma)$, and the matrices in $\U(\sigma)$ commute with $P(\sigma)$.

\section{Geometric phase}
\subsection{Geometric phase for unitarily evolving mixed states}
Inspired by the classical work of  Pancharatnam \cite{Pancharatnam1956},
Samuel and Bhandari \cite{Samuel_etal1988} generalized Berry's phase to a relative geometric phase for non-cyclic unitary evolutions of pure quantum states. 
Here we extend \eqref{phase from holonomy} to a relative geometric phase for non-cyclic unitary evolutions of mixed quantum states.

Let $\rho$ be a curve in $\D(\sigma)$.
The horizontal lifts of $\rho$ define
a parallel transport operator $\Pi[\rho]$ from the fiber over $\rho(0)$ onto the fiber over $\rho(\tau)$ according to 
$\Pi[\rho]\Psi_0=\Psi_{||}(\tau)$, where $\Psi_{||}$ is the horizontal lift of $\rho$ extending from $\Psi_0$. 
We define the \emph{geometric phase factor} and \emph{geometric phase} of $\rho$ to be
\begin{align}
&\geopf(\rho)=\tr\left(\Psi_{0}^\dagger\Pi[\rho]\Psi_{0}\right),\label{geophasefactor}\\
&\geop(\rho)=\arg\geopf(\rho),\label{geophase}
\end{align}
respectively. Observe that the geometric phase factor does not depend on $\Psi_0$ since the right hand side of \eqref{geophasefactor} is invariant under the transitive action of the gauge group on the fiber over $\rho(0)$. 
Definition \eqref{geophase} reduces to \eqref{phase from holonomy} if the curve is closed. However, for this definition to be useful in practice, we need a way to derive a horizontal lift from an arbitrary lift of a curve in $\D(\sigma)$. For this we need a connection form. 

\begin{figure}[htbp]
\begin{center}
\includegraphics[width=0.75\textwidth]{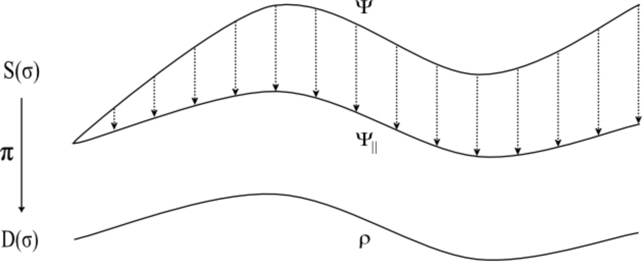}
\caption{Action of the exponential factor in Equation \eqref{horiz}.}
\label{figure}
\end{center}
\end{figure}

The Lie algebra of the gauge group is $\u(\sigma)=\left\{\xi\in\u(n):\xi P(\sigma)=P(\sigma)\xi\right\}$, and the infinitesimal generators of the gauge group yield canonical isomorphisms between $\u(\sigma)$ and the fibers in $\V\S(\sigma)$: $\u(\sigma)\ni\xi\mapsto \Psi\xi\in\V_\Psi\S(\sigma)$.
Furthermore, the horizontal bundle is the kernel bundle of the  \emph{mechanical connection form}
$\A:\T\S(\sigma)\to\u(\sigma)$ defined by $\A_{\Psi}=\I_{\Psi}^{-1}\J_{\Psi}$,
where $\I:\S(\sigma)\times\u(\sigma)\to \u(\sigma)^*$ and $\J:\T{\S(\sigma)}\to \u(\sigma)^*$ are the \emph{locked inertia tensor} and \emph{moment map}, respectively:
\begin{equation*}\label{tensors}
\I_{\Psi}\xi\cdot \eta=G(\Psi\xi,\Psi\eta),
\qquad
\J_{\Psi}(X)\cdot\eta=G(X,\Psi\eta).
\end{equation*}
If $\rho$ is a curve in $\D(\sigma)$, and $\Psi$ is any lift of $\rho$ to $\S(\sigma)$, then the horizontal lift of $\rho$ that extends from $\Psi(0)$ is 
\begin{equation}\label{horiz}
\Psi_{||}(t)=\Psi(t)\ptexp\left(-\int_{0}^{t}\A_\Psi(\dot \Psi)\dt\right),
\end{equation}
where $\ptexp$ is the positive time-ordered exponential.
The exponential factor in \eqref{horiz} has a nice geometric interpretation; it describes a curve in the gauge group 
$\U(\sigma)$ whose continuous action on $\S(\sigma)$ pushes $\Psi$ onto a horizontal curve, see Figure \ref{figure}.

Equations \eqref{geophase} and \eqref{horiz} give a truly general definition of geometric phase for unitary evolutions of mixed states. We will now derive an explicit formula for the geometric phase, suitable for computations. Thus, let 
$m_1,m_2,\dots m_l$ be the multiplicities of the \emph{distinct} eigenvalues in $\sigma$, where $m_1$ is the multiplicity of the greatest eigenvalue, $m_2$ of the second greatest eigenvalue and so forth, and define $l$ diagonal matrices by 
\begin{equation*}
E_k=\diag\left(\0_{m_1},\dots,\0_{m_{k-1}},\1_{m_k},\0_{m_{k+1}},\dots,\0_{m_l}\right).
\end{equation*} 
A straightforward computation shows that 
$\I_\Psi(\sum_k E_k\Psi^\dagger XE_k P(\sigma)^{-1})\cdot\eta=\J_\Psi(X)\cdot\eta$  
for every vector $X$ tangent to $\S(\sigma)$ at $\Psi$ and $\eta$ in $\u(\sigma)$. Hence, 
\begin{equation}\label{explicit}
\A_\Psi(X)=\sum_k E_k\Psi^\dagger XE_kP(\sigma)^{-1}.
\end{equation}
Moreover, if $U$ is a time-evolution operator that generates $\rho$, i.e., $\rho(t)=U(t)\rho(0) U(t)^\dagger$,
we can take $\Psi=U\Psi_0$, where $\Psi_0$ is any element in the fiber over $\rho(0)$. According to \eqref{geophase},\eqref{horiz}, and \eqref{explicit},
\begin{equation*}
\geop(\rho)=\arg\tr\left(\Psi^\dagger_0U(\tau)\Psi_0 \ptexp\left(-\int_{0}^{\tau}\sum_k E_k\Psi_0^\dagger U^\dagger(t)\dot U(t)\Psi_0 E_kP(\sigma)^{-1}\dt\right)\right).
\end{equation*}

\subsection{The geometric phase of Sj\"{o}qvist \emph{et al.}}\label{sjoqvist}
Generically, a density operator of a mixed state is nondegenerate. In \cite{Sjoqvist_etal2000}, the authors
gave a definition of geometric phase for \emph{parallel transported} nondegenerate mixed states.
We will quickly review it here, and then show that it is a special case of our definition.

Let the curve $\rho=\sum_k p_k\ketbra{k}{k}$ represent a unitary evolution of a non\-degenerate density operator.
Following Sj\"{o}qvist \emph{et al.} we say that the initial density operator $\rho(0)$ is parallel transported provided all of its eigenkets are parallel transported: $\braket{k}{\dot k}=0$.
The geometric phase of $\rho$ in the sense of \cite{Sjoqvist_etal2000} is then the argument of the expected value of the relative
phase factors for the eigenket curves:
\begin{equation}\label{sjoqvist phase}
\geop(\rho)=\arg\Big(\sum_k p_k\braket{k(0)}{k(\tau)}\Big).
\end{equation}

To show that \eqref{geophase} generalizes \eqref{sjoqvist phase}, let $\rho$ be any curve in $\D(\sigma)$, $\Psi_{||}$ be any horizontal lift of $\rho$, and set $\Psi_0=\Psi_{||}(0)$.
Define curves of normalized, and at each instant pairwise orthogonal,  kets 
$\ket{k}=p_k^{-1/2}\Psi_{||} e_k$. Then 
$\rho=\sum_{k} p_k\ketbra{k}{k}$,
and it follows from the horizontality of $\Psi_{||}$, and the observation
$\braket{k}{\dot k}=e_k^\dagger\A_{\Psi_{||}}(\dot\Psi_{||})e_k$,
that each curve of eigenkets is parallel. (In fact, using \eqref{explicit}, one can show that the horizontality condition $\A_{\Psi_{||}}(\dot\Psi_{||})=0$ is equivalent to the statement that, at each instant $t$, the velocity vector $\ket{\dot k(t)}$ is orthogonal to the eigenspace of $\rho(t)$ corresponding to the eigenvalue $p_k$.)
Now
\begin{equation*}
\tr\left(\Psi_0^\dagger\Pi[\rho]\Psi_0\right)
=\sum_k e_k^\dagger \Psi_{||}^\dagger(0)\Psi_{||}(\tau) e_k
=\sum_k p_k\braket{k(0)}{k(\tau)}.
\end{equation*}
This shows that our definition of geometric phase generalizes the definition of geometric phase in \cite{Sjoqvist_etal2000}.

\subsection{A mixture of neutrons in a magnetic field I}\label{neutron}
Consider an ensemble of neutrons subjected to an external magnetic field, and
assume that the ensemble is prepared so that the proportion of neutrons with spin up polarization is $p_1$ and the proportion with spin down polarization is $p_2$, where $p_1\ne p_2$.
The initial state of the spin part  of the ensemble's wave function can be represented by the density operator
\begin{equation*}
\rho_0=\begin{bmatrix} 
p_1 & 0\\ 
0 & p_2
\end{bmatrix}.
\end{equation*}
(Here we have modeled the system on $\CC^2$ in such a way that $e_1$ and $e_2$ represent the spin up and spin down states, respectively.)
If we further assume that the magnetic field is static and uniform with direction $\mathbf{n}=(\sin\theta,0,\cos\theta)^T$, then the Hamiltonian is given by
$H=-\omega\mathbf{n}\cdot\boldsymbol\sigma$, where $\omega$ depends on the magnitudes of the magnetic field and of the neutrons' magnetic moments, and
$\boldsymbol\sigma$ is the vector of Pauli matrices.
The time-evolution operator associated with $H$ is the one-parameter family of unitaries
\begin{equation*}
U(t)=\cos(\omega t)\begin{bmatrix} 
1 & 0\\ 
0 & 1
\end{bmatrix}+i\sin(\omega t)
\begin{bmatrix} 
\cos\theta & \sin\theta\\ 
\sin\theta & -\cos\theta
\end{bmatrix}.
\end{equation*}
Observe that this family is non-parallel transporting if $\cos\theta\ne 0$.
Now, a lift of the ensemble's evolution curve $\rho=U\rho_0U^\dagger$ is given by $\Psi=U\Psi_0$, where
\begin{equation*}
\Psi_0=\begin{bmatrix} 
\sqrt{p_1} & 0\\ 
0 & \sqrt{p_2}
\end{bmatrix}.
\end{equation*}
If the final time equals $\pi/\omega$, the curve $\rho$ returns to its initial state, and the geometric phase of $\rho$ is 
\begin{equation*}
\begin{split}
\geop(\rho)
&=\arg\tr\Big(\Psi^\dagger_0U(\pi/\omega)\Psi_0 \ptexp\Big(-\int_0^{\pi/\omega}\A_{\Psi}(\dot\Psi)\dt\Big)\Big)\\
&=\arg\left(p_1 e^{i\pi(1+\cos\theta)}+p_2 e^{i\pi(1-\cos\theta)}\right).
\end{split}
\end{equation*}

\section{Higher order geometric phases}
If the initial and final purifications $\Psi_0$ and $\Pi[\rho]\Psi_0$ are orthogonal, the geometric phase factor 
\eqref{geophasefactor} vanishes, and, consequently, the geometric phase \eqref{geophase} is not defined.
In this section we show how submatrices of $\Psi_0^\dagger\Pi[\rho]\Psi_0$ can be combined into higher order 
geometric phase factors that might be nonzero even if the  
geometric phase factor is zero. For a cyclically evolving state these phase factors will be formal homogeneous polynomials 
in the matrix entries of the holonomy of the evolution curve (hence the name 'higher order' geometric phase factors).
We also formulate a fairly general condition that guarantees the existence of a well-defined geometric phase of some order
not greater than the length of $\sigma$, and we observe that this condition is satisfied by cyclic evolutions.

\subsection{Geometric phase of degree $d$.}
Suppose $\rho$ is a curve in $\D(\sigma)$, 
and $\Psi_0$ is an element in the fiber over $\rho(0)$.
For each sequence $J=(j_1,j_2,\dots,j_d)$ of elements from $\{1,2,\dots,l\}$ we define the 
\emph{geometric phase factor and geometric phase for $\rho$ corresponding to $J$} by
\begin{align*}
\zeta_J(\rho)=\tr\prod_{k=1}^d E_{j_k}\Psi_0^\dagger\Pi[\rho]\Psi_0E_{j_{k+1}}\quad (j_{d+1}=j_1),\qquad\gamma_J(\rho)=\arg\zeta_J(\rho),
\end{align*}
respectively. The geometric phase factor \eqref{geophasefactor} can be recovered from the geometric phase factors corresponding to single element sequences:
$\geopf(\rho)=\sum_{j=1}^l\zeta_{(j)}(\rho)$.
More generally, if $d$ is any positive integer, then $\tr((\Psi_0\Pi[\rho]\Psi_0)^d)
=\sum_J\zeta_J(\rho)$,
where the sum is over all possible sequences $J=(j_1,j_2,\dots,j_d)$ that can be formed from the elements in $\{1,2,\dots,l\}$.
We define the \emph{$d^\th$ order geometric phase factor and geometric phase of $\rho$} as
\begin{equation*} 
\geopf^d(\rho)=\sum_J\zeta_J(\rho),\qquad \geop^d(\rho)=\arg\geopf^d(\rho).
\end{equation*}
Below we show that there is always a nonzero geometric phase factor of an order not greater than the length of $\sigma$ if $\Psi_0^\dagger\Pi[\rho]\Psi_0$ has a nonzero eigenvalue.

For cyclic evolutions the geometric phase factor $\zeta_J(\rho)$ vanishes if the sequence $J$ is not constant.
Indeed,
\begin{equation*}
\prod_{k=1}^d E_{j_k}\Psi_0^\dagger\Pi[\rho]\Psi_0E_{j_{k+1}}
=\begin{cases}
P(\sigma)^d\Hol(\rho,\Psi_0)^d E_{j_1}\text{if $j_1=j_2=\dots=j_d$},\\
\0_n\text{ otherwise,}
\end{cases}
\end{equation*}
due to the block diagonal structure of the matrices in $\U(\sigma)$.
It is apparent from this equation that $\zeta_{J}(\rho)$ is a formal 
homogeneous polynomial of degree $d$ in the entries of $\Hol(\rho,\Psi_0)$. 
Moreover, the matrix $\Psi_0^\dagger\Pi[\rho]\Psi_0=P(\sigma)\Hol(\rho,\Psi_0)$ has a nonzero eigenvalue because 
it represents an isomorphism of $\CC^n$. Thus, a cyclic $\rho$ has a well-defined geometric phase of some order.
(On the other hand, if $\rho$ is open and the initial and final states $\rho(0)$ and $\rho(\tau)$ are \emph{distinguishable} \cite{Englert1996,Markham_etal2008}, then
$\Psi_{0}^{\dagger}\Pi[\rho]\Psi_{0}=\0_{n}$ why $\zeta_{J}(\rho)=0$ for every sequence $J$, c.f. \cite{ED}.)

\subsection{A mixture of neutrons in a magnetic field II}
Consider the ensemble of neutrons in section \ref{neutron}, but now assume that $H$ is such that 
\begin{equation*}
U(t)=
\begin{bmatrix}
\cos(\omega t) & \sin(\omega t)\\
\sin(\omega t) & -\cos(\omega t)
\end{bmatrix}.
\end{equation*}
If the final time is $\pi/2\omega$, then $\geopf(\rho)=0$ and $\geopf^2(\rho)=2p_1p_2$.
Consequently, the geometric phase of $\rho$ is \emph{not} defined, but its second order geometric phase \emph{is} defined. 

\subsection{The existence of higher order geometric phases}
If $\Psi_0^\dagger\Pi[\rho]\Psi_0$ has at least one nonzero eigenvalue, then 
$\rho$ has a well-defined geometric phase of some order less than or equal to the length of $\sigma$.
To see this observe that the characteristic polynomial of $\Psi_0^\dagger\Pi[\rho]\Psi_0$ can be written 
\begin{equation*}
\det\big(\lambda-\Psi_0^\dagger\Pi[\rho]\Psi_0\big)=\sum_{k=0}^{n}(-1)^ks_k\lambda^{n-k},
\end{equation*}
where
\begin{equation*}
s_0=1,\quad s_k=\frac{1}{k}\sum_{j=1}^{k}(-1)^{j-1}s_{k-j}\geopf^j(\rho).
\end{equation*}
Therefore $\geopf^d(\rho)\ne 0$ for some $d\leq n$ if $\det\big(\lambda-\Psi_0^\dagger\Pi[\rho]\Psi_0\big)=0$ for some nonzero $\lambda$.

\subsection{Relations to the off-diagonal geometric phase factors of Manini and Pistolesi}
It is customary to represent an evolving density operator as an incoherent sum of 1-dimensional
time-dependent projectors defined by the normalized kets in a comoving eigenframe: $\rho(t)=\sum_k p_k\ketbra{k(t)}{k(t)}$.
This representation is unique for nondegenerate mixed states, and if the eigenkets are parallel transported, the geometric phase factor is the expected value of the relative phase factors $\braket{k(0)}{k(\tau)}$, see section \ref{sjoqvist}.

If all the relative phase factors vanish, the geometric phase is not defined.
Manini and Pistolesi \cite{Manini_etal2000} then proposed to study the gauge invariant \emph{off-diagonal phase factors}
\begin{equation*}\label{offisar}
\gamma_{k_1,k_2,\dots,k_d}^{(d)}(\rho)=\prod_{j=1}^d \braket{k_j(0)}{k_{j+1}(\tau)}\quad (\ket{k_{d+1}}=\ket{k_1}).
\end{equation*} 
Mukunda \emph{et al.} \cite{Mukunda_etal2001} extended the work of Manini and Pistolesi. 
They expressed the off-diagonal phase factors in terms of Bargmann invariants, and used the fact that
the Bargmann invariants satisfy certain relations, to give a full description of how the off-diagonal phase
factors are interrelated.

For degenerate mixed states the off-diagonal phase factors need not be gauge invariant.
However, appropriate combinations of them are. For example,
\begin{equation}\label{exp}
\zeta_{(j_1,j_2,\dots,j_d)}(\rho)=\sum_{k_1}\sum_{k_2}\dots\sum_{k_d}p_{k_1}p_{k_2}\dots p_{k_d}\gamma_{k_1,k_2,\dots,k_d}^{(d)}(\rho),
\end{equation}
where each index $k_s$ run through all integers satisfying
$0<k_s-\sum_{i=1}^{j_s-1}m_i\leq m_{j_s}$.
Hence, using \eqref{exp} and the results of Mukunda \emph{et al.}, one can, in principle, describe 
all relations between the higher order geometric phases. 

\section{Conclusion}
We have presented a fiber bundle framework in which geometric phase for mixed quantum states has a natural definition. This definition extends the standard definition of geometric phase of Sj\"oqvist \emph{et al.} \cite{Sjoqvist_etal2000}, which is based on interferometry, and is particularly well suited for computations in quantum physics. Furthermore, we have shown that the geometric phase is the first in a sequence of higher order geometric phases, and we have formulated a criterion that guarantees the existence of a geometric phase of some order. This criterion is met by cyclic evolutions of mixed states. 

Another approach to geometric phase for mixed states is given in Uhlmann \cite{Uhlmann1986, Uhlmann1991}. 
Uhlmann's approach only applies to mixed states represented by invertible density operators, and according to Slater \cite{Slater2002} it generally yields different outcomes than that of Sj\"{o}qvist \emph{et al.}, and hence ours. The discrepancy is probably due to a difference in the notions of parallel transport \cite{Filipp_etal2005, ED}. 
Also Singh \emph{et al.} \cite{Singh_etal2003} give a definition of geometric phase for parallel transported mixed states\footnote{Unfortunately, the proof of gauge invariance of the proposed geometric phase in \cite{Singh_etal2003} lacks vital parts. The authors are forced to introduce certain initial values in the proof that has as a consequence that the proof does not imply full gauge invariance. Moreover, the definition of geometric phase is such that 
for an evolution of a noninvertible density operator, the holonomy of an artificially chosen frame of kets in the eigenspace corresponding to the density operator's zero eigenvalue seems to contribute to the geometric phase. We don't think this is correct from a physical point of view.
Also, the definition does not, in an obvious way, reduce to the Aharonov-Anandan geometric phase for pure states, and it is not clear how to generalize the definition to infinite dimensional Hilbert spaces.}, and the paper by Chaturvedi \emph{et al.} \cite{Chaturvedi_etal2004} contains a differential geometrical approach to geometric phase for mixed states.
Comparative studies of our definition and those in \cite{Singh_etal2003} and \cite{Chaturvedi_etal2004} will be published in forthcoming papers.

Furthermore, in a forthcoming paper, off-diagonal geometric phases for mixed states will be defined and investigated.
Such phases have been studied by Filipp and Sj\"{o}qvist \cite{Filipp_etal2003a, Filipp_etal2003b, Filipp_etal2005}.
In \cite{Filipp_etal2005}, the authors defined ``off-diagonal quantum holonomy invariants'' for the bundle of purifications by Uhlmann, and compared these with the off-diagonal geometric phase factors defined in \cite{Filipp_etal2003b}. 
The conclusion was that there is ``a general discrepancy for these two approaches related to a fundamental difference in the
treatment of parallel transport of quantum states''.
The off-diagonal quantum holonomy invariants can be similarly defined for the bundle of purifications in the current paper, and we will investigate how these are related to the off-diagonal geometric phases by Filipp and Sj\"{o}qvist.

\section*{Acknowledgement}
The second author acknowledges the financial support by the Swedish Research Council (VR), grant number 2008-5227.

\end{document}